\documentclass[journal=jctcce,manuscript=article,layout=traditional, chaptertitle=true]{achemso}
\pdfoutput=1

\ifpdf
  \pdftrailerid{}
\fi

\usepackage{amssymb}
\usepackage[T1]{fontenc} % Use modern font encodings
\usepackage{mathtools}
\usepackage{graphicx}
\usepackage{multirow}
\usepackage{siunitx}
\usepackage{hyperref}
\usepackage{booktabs}
\usepackage{pdfpages}

\SectionNumbersOn
\setkeys{acs}{doi = true}
\newcommand{\doi}[1]{\href{https://doi.org/#1}{\nolinkurl{#1}}}
\title{
    Reliable Viscosity Calculation from High-Pressure Equilibrium Molecular Dynamics: Case Study of 2,2,4-Trimethylhexane
}

\author{G\"{o}zdenur Toraman}
\affiliation[Soete]{Soete Laboratory, Ghent University, Technologiepark-Zwijnaarde 46, 9052 Ghent, Belgium}
\author{Dieter Fauconnier}
\affiliation[Soete]{Soete Laboratory, Ghent University, Technologiepark-Zwijnaarde 46, 9052 Ghent, Belgium}
\alsoaffiliation[FlandersMake]{FlandersMake@UGent, Core Lab MIRO, 3001 Leuven, Belgium}
\author{Toon Verstraelen}
\affiliation[Ghent University]{Center for Molecular Modeling (CMM), Ghent University, Technologiepark-Zwijnaarde 46, B-9052, Ghent, Belgium}
\email{toon.verstraelen@ugent.be}

\begin{document}

%\begin{center}
%    Version
%    \input{../gitline.txt}
%\end{center}

\newpage

\begin{abstract}
    Viscosity is a fundamental property of liquid lubricants, yet it is challenging to determine accurately, especially at high pressures.
    Although equilibrium molecular dynamics (EMD) simulations are a promising alternative to resource-intensive experiments, practical challenges remain in assessing the sufficiency of simulation time and in controlling uncertainties in the Green--Kubo formalism due to the finite amount of trajectory data.
    In this work, we extend the STable AutoCorrelation Integral Estimator (STACIE), a recently developed algorithm for estimating transport properties.
    First, we introduce the Lorentz model to estimate the viscosity and the exponential correlation time from the low-frequency power spectrum of deviatoric pressure fluctuations.
    Second, we show how to supplement the three conventional off-diagonal elements of the pressure tensor ($P_{xy}$, $P_{yz}$ and $P_{zx}$) with two additional uncorrelated deviatoric pressure components for shear viscosity calculations.
    Using these improvements, we apply STACIE to calculate the shear viscosity of \textit{2,2,4-trimethylhexane} from EMD simulations.
    We demonstrate STACIE's capability to reliably calculate viscosity under high-pressure conditions, offering a robust and automated solution with validated uncertainty quantification.
    Our results, when compared to the outcomes of the 10$^\text{th}$ International Fluid Properties Simulation Challenge (IFPSC), underscore the need for long EMD simulations.
    Large deviations from experimental viscosities in previous works were primarily due to insufficient simulation times and \textit{ad hoc} post-processing choices, rather than the limitations of the force fields used.
    Unlike previous studies, our viscosity estimates agree well with experimental results (relative error < 4\%) up to the highest pressure of \SI{1}{\giga\pascal}, highlighting the improved reliability and accuracy of STACIE's systematic approach to viscosity predictions.
\end{abstract}

\begin{tocentry}
    \includegraphics{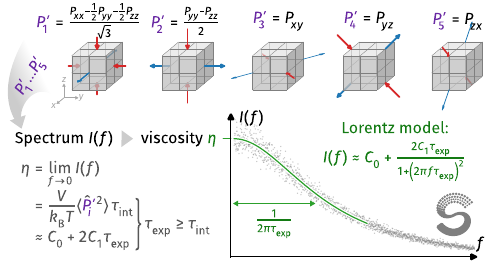}
\end{tocentry}

\newpage

\section{Introduction}\label{sec:introduction}
% General relevance of viscosity.
Shear viscosity is a fundamental transport property of fluids, both liquids and gases, that quantifies their resistance to shear flow as a result of momentum exchanged between nearby fluid molecules.
Accurate viscosity input is essential for modeling fluid dynamics across scientific and engineering disciplines, including aeronautics, medicine, geophysics, astrophysics, high-energy physics, polymer chemistry and pharmaceutics.\cite{Connes2013, Chevrel2019, Wesp2011, Deng2024, Ofengeim2015, Bishop2022, Dobrynin2023, Janchai2023, Brubaker2024}
In tribology, a sub-discipline of mechanical engineering, accurate characterization of the viscosity of lubricants in a relevant range of pressures, shear rates and temperatures, holds paramount importance.
Such knowledge is crucial for selecting a suitable and optimal lubricant for machines and their mechanical components (gears, bearings, \ldots) as it affects lubricant film thickness ($\approx\SI{1}{\micro\meter}$) between solid surfaces, their energy efficiency and lifetime.

% Challenges of viscosity measurements and calculations at high-pressures.
Unfortunately, dependable, precise, and complete data on viscosity for different lubricants remains scarce.
In particular for high pressures (> \SI{1}{\giga\pascal}), relevant for heavily loaded contacts operating under elastohydrodynamic lubrication (EHL) and thermo-elastohydrodynamic lubrication (TEHL) conditions, experimental data is very limited.
Performing ultra-high-pressure experiments, even up to \SI{1}{\giga\pascal}, is notably challenging and requires highly specialized equipment.\cite{Bair2019High}
Since the early 2000s, Molecular Dynamics (MD) has gained traction as a promising alternative.\cite{Daivis1994, McCabe2001, Martini2006}
With only the molecular structure and an appropriate force field as inputs, MD simulations can, in principle, make predictions and give detailed atomistic insight at any thermodynamic state.\cite{Ewen2018, Spikes2014, Maginn2020}
However, accurate predictions of bulk and shear viscosity of hydrocarbon lubricants using MD are notoriously challenging, even at ambient conditions.
These challenges exacerbate under very high pressures, since the viscosity of hydrocarbon lubricants is known to increase nearly exponentially with pressure,\cite{Bair2019High} significantly slowing down molecular motion and complicating statistical sampling from simulations.

% Introduce MD methods and their challenges. Situate this work.
In MD, shear viscosity can be determined using either EMD or non-equilibrium MD (NEMD) methods, each providing complementary information depending on the target rheological regime.\cite{Maginn2020}
NEMD is particularly powerful for characterizing non-Newtonian rheology and shear-thinning behavior under high shear rates,\cite{Todd2007} which are directly relevant for many applications, including EHL.\cite{Ewen2021}
In contrast, EMD provides direct access to the zero-shear viscosity from equilibrium fluctuations and can yield multiple transport properties from the same simulation.\cite{Chen2009}
However, reliable viscosity estimates from equilibrium molecular dynamics (EMD) simulations of lubricants at high pressures are notoriously difficult to obtain due to the slow convergence of off-diagonal pressure autocorrelation functions.
In this work, we extend STACIE, a general algorithm for transport property calculations,\cite{Toraman2025} to facilitate challenging zero-shear viscosity estimates.

% Recap of Green--Kubo theory.
Green--Kubo theory is the main theoretical framework to derive transport properties from EMD simulations.\cite{Kubo1957, Green1954}
For example, it relates shear viscosity to the autocorrelation of microscopic off-diagonal pressure tensor fluctuations.\cite{Daivis1994, Allen2017, Frenkel2002}
This should not be confused with empirical relations that estimate shear viscosity from diffusion coefficients\cite{Sengupta2013, Puosi2019} or from their system-size dependence,\cite{Yeh2004, Jamali2018} which can also be derived from EMD simulations.
Using the notation introduced in our previous work,\cite{Toraman2025} shear viscosity can be expressed as follows:
\begin{align}
    \eta = \frac{V}{2 k_\text{B} T} \int_{-\infty}^{+\infty} \underbrace{\Bigl\langle\hat{P}_{xy}(t_0) \hat{P}_{xy}(t_0 + \Delta_t)\Bigr\rangle}_{\text{ACF, } c(\Delta_t)} \mathrm{d}\Delta_t
    \label{eq:gk}
\end{align}
where $V$ is the volume of the simulation cell, $k_\text{B}$ is the Boltzmann constant, $T$ is the temperature,
and $\hat{P}_{xy}$ is an off-diagonal element of the pressure tensor.
As will be discussed in this work, the statistical convergence can be improved by averaging the viscosity over five uncorrelated deviatoric pressure components.\cite{Holian1983, Daivis1994, Alfe1998}
The integrand in Eq.~\eqref{eq:gk} is the autocorrelation function (ACF) of $\hat{P}_{xy}$, denoted as $c(\Delta_t)$, where $\Delta_t$ is the time lag.

% Overview of algorithms (indirectly) based on Green-Kubo.
Various algorithms have been developed to estimate transport properties from finite amounts of EMD trajectory data, all of which can be justified from Green--Kubo theory.
As outlined in our previous work,\cite{Toraman2025} they can be broadly classified into four categories:
Green--Kubo (GK) algorithms, which truncate the integral in Eq.~\eqref{eq:gk} and use a sampling estimate of the ACF;\cite{Zhang2015, OteroLema2025, Meel2026}
Einstein--Helfand (EH) algorithms based on mean squared displacements;\cite{Helfand1960, Pranami2015}
cepstral\cite{Ercole2017, Ercole2022, Drigo2024, Wieser2025} and spectral methods.\cite{Pegolo2025, Toraman2025}
The latter two categories utilize the Wiener--Khinchin theorem to relate the transport property to the zero-frequency limit of (the logarithm of) a power spectrum.

% Zoom in on TDM as a popular but problematic GK method.
The time decomposition method (TDM) proposed by Zhang \textit{et al.}\cite{Zhang2015} falls into the category of GK algorithms,
and it has been widely used for shear viscosity estimation of lubricants at elevated pressures.\cite{Kondratyuk2019, Messerly2019, Ewen2021, Kruse2024}
Although the TDM achieved promising results in some cases, it relies on \textit{ad hoc} choices and human judgment, which limits its robustness and reproducibility.
Originally developed and tested for systems with relatively fast dynamics, its applicability to high-pressure systems is uncertain.
In fact, as demonstrated in our previous work,\cite{Toraman2023} the visual assessment of intermediate plots required in TDM can severely bias the predicted viscosity in more challenging cases.
Higher pressures exacerbate these difficulties because viscosity increases and pressure fluctuations relax more slowly.

% Zoom out again: project the symptoms of TDM as a broader issue in EMD methods.
This reveals a broader issue:
Many commonly used methods for evaluating transport properties have primarily been tested under ambient or high-temperature conditions, or on relatively simple systems
(typically Lennard-Jones fluids) where system dynamics are fast (lower viscosities) and correlation times are short.
However, under high-pressure conditions, with high viscosity and long correlation times, it becomes far more difficult to achieve reliable estimates of transport properties.
These challenges become especially pronounced in the (T)EHL regime and raise concerns about the reliability of EMD results,
as evidenced in the findings of the 10$^\text{th}$ IFPSC.\cite{Kondratyuk2019, Zheng2019, Messerly2019}

% Introduce STACIE as a solution to these broader issues.
To address these challenges, we recently developed STACIE (STable AutoCorrelation Integral Estimator),\cite{Toraman2025}
a spectral algorithm designed to tackle three major issues in EMD-based transport property estimation:
(i) the need for a robust, automated approach that avoids subjective human intervention in the analysis of EMD data;
(ii) the difficulty of determining the required simulation time, especially in systems with slow dynamics; and
(iii) the need for uncertainty quantification of the estimated transport properties for reliable and accurate estimates.
STACIE estimates a transport property by evaluating the zero-frequency limit of a model fitted to the low-frequency part of the power spectrum of the corresponding time-dependent quantity, which allows for a precise uncertainty quantification in a fully automated manner.

% Relation to hydrodynamics.
STACIE belongs to the class of spectral methods, which are relatively new and emphasize a rigorous statistical framework for uncertainty quantification of transport properties.\cite{Pegolo2025, Toraman2025}
However, the definition of a transport property as the zero frequency and wavevector limit of a spectrum has a rich history in the context of hydrodynamics.\cite{Maxwell1867, Evans1981, Evans1990, Hansen2013, Allen2017}
In modern MD studies of nanofluidics, the frequency and wavevector-dependent viscosity, $\eta(\omega, \mathbf{k})$, is used to assess deviations at the molecular scale from an idealized Newtonian fluid.\cite{Puscasu2010, Levashov2014, Xian2019}
While STACIE was originally developed as a general robust method for estimating the integral of the ACF, its use in viscosity estimation can be seen as a practical application of hydrodynamics.

% Objectives of this work.
The primary objective of this work is to demonstrate the applicability and performance of STACIE to calculations of the shear viscosity of \textit{2,2,4-trimethylhexane} up to a pressure of \SI{1}{\giga\pascal}.
To this end, we extend STACIE by introducing the ``Lorentz'' model for the low-frequency region of the power spectrum, which is specifically tailored for ACFs with exponentially decaying tails.
The secondary objective is to provide a physically grounded guideline for determining the required simulation time.
The exponential correlation time,\cite{Sokal1997} $\tau_\text{exp}$, is one of the parameters of the Lorentz model and corresponds to the width of the Lorentzian peak centered at frequency zero in the power spectrum.
Unlike the integrated correlation time, $\tau_\text{int}$, which is related to the error of the mean of time-correlated data,
$\tau_\text{exp}$ characterizes the slowest relaxation in the system.
We demonstrate how $\tau_\text{exp}$ can be used to assess whether an MD simulation is sufficiently long to capture these slowest modes.
Our new approach offers a more systematic alternative to existing \textit{ad hoc} guidelines for determining the simulation time.\cite{Carlson2022, Maginn2020}
We compare our results with those reported in the 10$^\text{th}$ IFPSC and discuss the implications for high-pressure viscosity calculations.

% Structure of the paper.
The remainder of the paper is structured as follows: Section~\ref{sec:model} introduces the Lorentz model, which is applicable to any autocorrelation function with exponentially decaying tails.
This section also provides practical guidelines for preparing inputs to ensure reliable estimates when applying this new model.
Section~\ref{sec:model} concludes with a discussion of how exponential and integrated correlation times differ and how they are related.
Section~\ref{sec:emd} starts with a brief description of the EMD dataset employed in this study.
It also shows how to incorporate five uncorrelated deviatoric pressure components,
as opposed to the usual three, to improve the statistical efficiency of viscosity estimates.
Section~\ref{sec:results} presents the viscosity results obtained with STACIE under both ambient and high-pressure conditions,
and includes comparisons with experimental data and previous MD predictions in the literature.
Additionally, it discusses the influence of simulation duration on the reliability of viscosity estimates.
This section concludes with a short numerical validation of the five uncorrelated deviatoric pressure components.
Finally, Section~\ref{sec:conclusions} summarizes the key findings and outlines potential directions for future work.
All error estimates are standard uncertainties unless they are specified as a 95\% confidence interval (CI).

\section{Power Spectrum Model}\label{sec:model}
% Section overview
This section begins with a brief overview of the core concept behind STACIE.
For a comprehensive explanation, including detailed mathematical derivations and implementation specifics,
readers are referred to our earlier work and STACIE's documentation.\cite{STACIE2025, Toraman2025}
We then introduce a model tailored for ACFs that exhibit exponential decay at long time lags, which is suitable for shear viscosity calculations.
With this new model, we revisit and extend our practical guidelines for preparing input data to ensure accurate and reliable viscosity estimates using STACIE.
Finally, we discuss the exponential and integrated correlation times, both of which can be estimated within this framework.

% Summary of STACIE and importance of the spectrum
In essence, STACIE constructs the sampling power spectral density (PSD),
fits a user-defined model to the low-frequency region of the spectrum, and evaluates the fitted model at zero frequency to obtain the autocorrelation integral,
$\mathcal{I}$, and its uncertainty, $\sigma_{\mathcal{I}}$.
The main advantage of spectral analysis is that the Fourier spectrum amplitudes have uncorrelated uncertainties at different frequencies, which simplifies uncertainty quantification.\cite{Ercole2017, Ercole2022, Drigo2024, Pegolo2025, Toraman2025}
To leverage this, STACIE employs locally weighted Bayesian regression to estimate parameters of the model fitted to the low-frequency part of the spectrum.
For numerical efficiency, the algorithm determines parameters that maximize the posterior probability and derives uncertainty estimates using the Laplace approximation.\cite{Toraman2025}
In contrast, ACFs and mean squared displacements have correlated uncertainties at different time lags,\cite{Bartlett1980, Francq2009, Jones2012, Kim2018} which are typically ignored in Green--Kubo and Einstein--Helfand algorithms,\cite{Pranami2015, Zhang2015, OteroLema2025, Meel2026} hampering a robust uncertainty quantification.

% Summarize the cutoff frequency, as we'll need this background to explain the penalty.
STACIE automatically detects a suitable low-frequency part of the spectrum defined by a cutoff frequency.
The robustness of STACIE's uncertainty quantification stems from considering a wide range of possible cutoff frequencies, $f_{\text{cut},j}$.
Instead of selecting a single best cutoff frequency, STACIE fits model parameters for all cutoffs, assigns a weight to each fit using cross validation, and marginalizes statistical estimates over the grid of frequency cutoffs.
This effectively accounts for the uncertainty associated with selecting the cutoff frequency.

% Overview of spectrum models in STACIE
The current version of STACIE (v1.2.1) supports three models for fitting the low-frequency region of the spectrum,\cite{STACIE2025}
namely the exponential polynomial model (exppoly), the Pad\'e model, and the Lorentz model.
Our earlier work introduced the exppoly model,\cite{Toraman2025} which represents the PSD as an exponential of a low-degree polynomial.
It offers broad applicability with minimal assumptions about spectral shape.
It also provides robust parameter estimation because initial guesses can be obtained through linear regression.
Its effectiveness was demonstrated using both synthetic data and the ionic conductivity of an electrolyte.
In the present work, we introduce a model suited for autocorrelation functions that decay exponentially,
characterized by a correlation time, $\tau_\text{exp}$, which represents the slowest relaxation mode of the system.
This parameter is particularly valuable for selecting appropriate simulation times and block sizes
for high-pressure systems with slow dynamics.
This extends the practical guidelines we previously outlined
for preparing sufficient input data.\cite{Toraman2025}

\subsection{The Lorentz model}\label{subsec:lorentz-model}

% Introduce and motivate the Lorentz model: better fit with few parameters than a generic model
Although STACIE's ExpPoly model is robust and widely applicable, the Lorentz model is specifically designed for input time series that exhibit exponentially decaying correlations.
In such cases, the Lorentz model can explain a larger portion of the low-frequency spectrum with three parameters compared to a generic model, which results in more robust fits.

% Motivate mathematical form of the model.
The Lorentz model assumes that the input time series exhibits one slow decay mode,
superimposed on significantly faster oscillations.
For large time lags, the ACF is assumed to have the following form:
\begin{align*}
    c(\Delta_t) \approx C_1 \, \exp\left(-\frac{| \Delta_t |}{\tau_\text{exp}}\right)
\end{align*}
where $C_1$ is the tail amplitude and $\tau_\text{exp}$ is the exponential correlation time.
If the ACF were purely exponential for all time lags,
the corresponding PSD would be a Lorentzian function centered at zero frequency:
\begin{align*}
    I^\text{pure-exp}(f) = \frac{2 C_1 \tau_\text{exp}}{1 + (2 \pi f \tau_\text{exp})^2}
\end{align*}
In practice, this pure exponential form is rarely observed because of intricate short-time dynamics.
If the short time dynamics occur on a timescale $\tau_\text{fast} \ll \tau_\text{exp}$,
the spectrum at low frequencies remains Lorentzian but with a white-noise background, represented by a constant $C_0$ in the final form of STACIE's Lorentz model:
\begin{align*}
    I^\text{lorentz}(f) = C_0 + \frac{2 C_1 \tau_\text{exp}}{1 + (2 \pi f \tau_\text{exp})^2}
    \quad \text{for } f \ll 1/\tau_\text{fast}
\end{align*}
To facilitate the regression and the uncertainty quantification,
STACIE internally represents this model as a special case of the Pad\'e model:
\begin{align*}
    I^\text{pade}(f) = \frac{
        \displaystyle
        \sum_{s \in S_\text{num}} p_s f^s
    }{
        \displaystyle
        1 + \sum_{s \in S_\text{den}} q_s f^s
    }
\end{align*}
with $S_\text{num} = \{0, 2\}$ and $S_\text{den} = \{2\}$.

% Remind the reader of the cutoff and explain the notation for different cutoffs.
As detailed in our previous work, STACIE fits a model to only the low-frequency spectrum, which is controlled by a cutoff frequency.
A logarithmic grid of cutoff frequencies is constructed and for each cutoff frequency, $f_{\text{cut},j}$, the regression is repeated.
To clarify this aspect, all fitted parameters are denoted with a superscript $(j)$ below.

% Explain the regression for a single cutoff frequency.
For a given cutoff $f_{\text{cut},j}$, the estimated Pad\'e model parameters $\{\hat{p}_0^{(j)}, \hat{p}_2^{(j)}, \hat{q}_2^{(j)}\}$
are transformed into Lorentz model parameters as follows:
\begin{align*}
  \begin{aligned}
    \hat{C}_0^{(j)} &= \frac{\hat{p}_2^{(j)}}{\hat{q}_2^{(j)}}
    \\
    \hat{C}_1^{(j)} &= \frac{\pi}{\sqrt{\hat{q}_2^{(j)}}}\left(\hat{p}_0^{(j)} - \frac{\hat{p}_2^{(j)}}{\hat{q}_2^{(j)}}\right)
    \\
    \hat{\tau}_\text{exp}^{(j)} &= \frac{\sqrt{\hat{q}_2^{(j)}}}{2 \pi}
  \end{aligned}
\end{align*}
Hats are used throughout this work to denote stochastic quantities.
The results for cutoff $j$ are only retained if the Pad\'e parameters correspond to a Lorentzian peak, namely if $\hat{q}_2^{(j)} > 0$ and $\hat{p}_0^{(j)}\, \hat{q}_2^{(j)} > \hat{p}_2^{(j)}$.
When these conditions are met,
STACIE performs the transformation to $\hat{C}_0^{(j)}$, $\hat{C}_1^{(j)}$, and $\hat{\tau}_\text{exp}^{(j)}$
and applies first-order uncertainty propagation to map the covariance of the Pad\'e parameters to the covariance of the Lorentz parameters.
By utilizing the Jacobian matrix of the transformation, the algorithm computes the full covariance matrix of the Lorentz parameters,
from which the uncertainties are derived as the square root of the diagonal elements.
This rigorous mapping allows STACIE to account for parameter correlations and report a robust uncertainty for the relaxation time $\hat{\tau}_\text{exp}^{(j)}$ and the resulting viscosity.
Finally, note that the estimate of the autocorrelation integral is simply $\hat{\mathcal{I}}^{(j)} = \hat{p}_0^{(j)}$.

% Heuristics for stable estimates of tau_exp, marginalization over cutoffs
To ensure stable estimates of $\hat{\tau}_\text{exp}$, STACIE applies two heuristics when marginalizing the fitted parameters over all frequency cutoffs.
First, fits with an extremely large relative uncertainty in $\hat{\tau}_\text{exp}^{(j)}$ are discarded, namely when
\begin{align*}
  \hat{\mathcal{R}}_j = \frac{
    \hat{\sigma}^{(j)}_{\tau_\text{exp}} / \hat{\tau}_\text{exp}^{(j)}
  }{
    \hat{\sigma}^{(j)}_{\mathcal{I}} / \hat{\mathcal{I}}^{(j)}
  } > 100.
\end{align*}
A large relative uncertainty in $\hat{\tau}_\text{exp}^{(j)}$ indicates that the cutoff frequency is so low that the model is fitted to the flat part of the maximum of the Lorentzian peak.
In this situation, we observed that the Laplace approximation for the parameter uncertainty tends to break down, leading to poor uncertainty estimates and an unreliable cross-validation.
Second, to further reduce the risk of including misleading fits, the weights used to marginalize over all cutoff frequencies also include an exponential penalty:
\begin{align*}
    \hat{W}_j^\text{lorentz} \propto \hat{\mathcal{L}}^\text{CV2L}_j
    \,\exp(-\hat{\mathcal{R}}_j)
    \,\operatorname{H}(100 - \hat{\mathcal{R}}_j)
\end{align*}
where $\operatorname{H}(\cdot)$ is the Heaviside step function, and $\hat{\mathcal{L}}^\text{CV2L}_j$ is a weight factor determined via cross-validation, as described in our original paper on STACIE.\cite{Toraman2025}
The exponential penalty gradually assigns lower weights to fits with larger relative uncertainties in $\hat{\tau}^{(j)}_\text{exp}$.
By expressing the penalty as a function of a ratio of relative uncertainties,
it becomes independent of the overall statistical quality of the input data.
This penalty mitigates unreliable fits and leads to a more robust estimate of $\hat{\tau}_\text{exp}$ after marginalization:
\begin{align*}
    \hat{\tau}_\text{exp} =
      \sum_j \hat{W}_j^\text{lorentz} \, \hat{\tau}_\text{exp}^{(j)}
    \quad \text{with} \quad
    \sum_j \hat{W}_j^\text{lorentz} = 1
\end{align*}
Final values of all fitted parameters are constructed using the same weighted average.

% Briefly mention validation with ACID test set
STACIE's implementation of the Lorentz model has been validated using the ACID test,
which also contains two test cases with exponentially decaying ACFs.\cite{Toraman2025, ACID2025}
This validation, summarized in Section~S1 of the Supporting Information,
confirms that both the autocorrelation integral and the exponential correlation time
are accurately estimated with reliable uncertainties.
One minor limitation is that the uncertainty in $\hat{\tau}_\text{exp}$ tends to be overestimated, by about 25\%.

% Applications and limitations
It may seem surprising that the Lorentz model can be applied to a wide range of systems,
despite its simplicity and the strong assumptions it makes about the ACF.\cite{Jones1992}
A well-known example that gives exactly this form of the ACF is the solution of a first-order Langevin equation,
which describes the velocity of a Brownian particle in a fluid.\cite{Kubo1991}
More complex systems often exhibit the same exponential decay at long time lags,
because their fast modes act as thermal noise perturbing the slowest degrees of freedom,
which is the essence of the first-order Langevin equation.
However, when there is no clear separation between slow and fast timescales,
the ACF may not exhibit an exponential tail, and the Lorentz model becomes less
suitable.\cite{Rowlands2008, Siegle2010}
STACIE's documentation includes several examples to illustrate
when the Lorentz model is appropriate and when it is not.\cite{STACIE2025}

% Discuss the expectations for the ACF of the deviatoric pressure fluctuations in viscosity calculations.
The applicability of the Lorentz model to shear viscosity calculations should not be taken for granted.
Early work on mode-coupling theory predicted a $\Delta_t^{-3/2}$ power-law decay of the ACF of the deviatoric pressure for a 3D isotropic fluid.\cite{Ernst1976, Alder1970}
This asymptotic behavior would imply a low-frequency spectrum of the form $a + b \sqrt{f}$\cite{Holian1983}, which is fundamentally incompatible with the Lorentz model.
These early models were later challenged, e.g., by extended mode coupling theory, which predicted a power law decay at intermediate time lags, followed by an exponential decay due to structural relaxation.\cite{Kirkpatrick1986, deSchepper1986}
While mode coupling theory has been successful in modeling the glass transition,
it employs a simplified model that does not necessarily capture the full atomistic detail of an MD simulation.
A systematic MD study of the density-dependence of the deviatoric pressure ACF of a monoatomic fluid
confirmed that power-law decay is only observed at intermediate time lags and high densities,
and that the ACF eventually decays exponentially.\cite{Hartkamp2013}
This is consistent with numerous other MD studies that have successfully fitted a sum of exponential functions to the tail of (the running integral of) the ACF.\cite{Hess2002, Fernandez2005, Basconi2013, Hartkamp2013, Zhang2015, Kondratyuk2019, Zheng2019, Kondratyuk2021, Odintsova2025, OteroLema2025, Meel2026}
Consequently, the Lorentz model is expected to have similarly broad applicability to EMD shear viscosity calculations.
In fact, without the white-noise term, the Lorentz model corresponds to the dissipative part of Maxwell's model for viscoelastic response, which has been successfully applied to model the frequency-dependent viscosity of fluids.\cite{Maxwell1867, Evans1990, Xian2019}
To accommodate the slow decay associated with the transient power-law regime of the ACF,
the stretched exponential (or molasses tail) model, $\propto \exp(-(t/\tau)^\beta)$ with $\beta < 1$, has been a popular choice.\cite{Isobe2008, Medina2010, Furukawa2011, Li2014, Guillaud2017a, Guillaud2017b, Meel2026}
This empirical model was originally proposed to study the dielectric relaxation of polymers.\cite{Williams1970}
Recently, Meel \textit{et al.} compared three different tail models,
namely exponential, stretched-exponential, and power-law decays,
and found that the exponential and stretched-exponential models yielded mutually consistent viscosity estimates,
whereas the power-law model systematically yielded significantly larger viscosities.\cite{Meel2026}
We have not yet implemented a spectral equivalent of the stretched exponential model in STACIE,
but we are considering it as a potential future extension.

\subsection{Guidelines for preparing sufficient inputs for the Lorentz model} \label{subsec:input-prep}

% Explain the circular dependency of sim time and block size
Accurate predictions of transport properties require MD simulations long enough
to capture the system's slowest modes.
These slow modes determine how the power spectral density (PSD) converges to its zero-frequency limit.
If the simulation time ($t_\text{sim}$) is too short,
the Discrete Fourier Transform (DFT) grid resolution ($1/t_\text{sim}$) becomes too coarse
to adequately resolve the low-frequency region of the spectrum,
which has two major consequences.
First, fitting a model to such low-resolution data results in a poor estimate of the autocorrelation integral and hence the value of the transport property.
Second, the exponential correlation time cannot be derived reliably from a low-resolution spectrum for the same reason.
Since this is an essential indicator for the simulation time required to obtain converged transport properties,
we emphasize that one can confirm only \textit{a posteriori} whether the chosen simulation time was sufficiently long.
In our previous work, we introduced some practical guidelines to systematically extend inputs until they pass STACIE's internal sanity checks, thereby offering a practical approach to obtaining sufficient data and accurate transport properties with reliable uncertainty estimates.\cite{Toraman2025}
More detailed information on these recommendations can be found in the STACIE documentation.\cite{STACIE2025}
Here, we briefly summarize them and present further guidelines tailored to the Lorentz model introduced in Section~\ref{subsec:lorentz-model},
by making use of the exponential correlation time, $\tau_\text{exp}$.

% Recommendations for ...
\begin{enumerate}
    % ... the number of independent sequences
    \item
    Running multiple independent simulations is essential for improving statistical accuracy.
    A single MD trajectory can yield multiple time-correlated input sequences, such as the three Cartesian components of the heat flux in thermal conductivity calculations.
    However, relying on only one trajectory, even if it provides several such sequences, typically results in poor statistical estimates, reflected in large error bars.
    Running multiple independent MD trajectories, each starting from distinct initial configurations or velocity distributions,
    enhances the reliability of the autocorrelation integral and leads to reduced statistical uncertainties.
    In our previous work, we recommended choosing the number of independent sequences as $M \approx 1/(20\,P\,\epsilon_\text{rel}^2)$, based on the target relative error, $\epsilon_\text{rel}$,
    of the autocorrelation integral and the number of model parameters, $P$.\cite{Toraman2025}
    We further recommended that each sequence should contain at least $N \geq 400\,P$ steps.

    % ... the number of effective points
    \item
    For a given cutoff frequency, the DFT grid points are assigned weights that are one for the lowest frequencies
    and that gradually decrease to zero around the cutoff frequency.
    The sum of these weights is the effective number of data points used in the fit, $N_\text{eff}$.
    As a general guideline, we recommend using at least $20\,P$ effective points.
    For the Lorentz model, this corresponds to a minimum of $N_\text{eff} \geq 60$
    to ensure that the peak shape is captured accurately.

    % .. simulation time
    \item
    As mentioned earlier, an ACF with exponentially decaying tails has a corresponding spectrum featuring a Lorentzian peak at zero frequency,
    whose peak width equals $1/(2 \pi \tau_\text{exp})$.
    To obtain a reliable fit, this peak must be well-resolved on the DFT frequency grid, whose spacing is $1/t_\text{sim}$.
    A sufficiently fine grid is achieved when the simulation time is much longer than $2\pi \tau_\text{exp}$.
    Accordingly, we recommend a minimum simulation time of:
    \begin{align}
       t_{\min} \approx 20\, \pi \tau_\text{exp}
        \label{eq:min-simtime}
    \end{align}
    This guideline ensures that the Lorentzian peak is well-resolved and the Lorentz model can accurately capture the slowest mode of the system.
    It is not intended as a strict requirement,
    but rather as another useful quantity to monitor when using STACIE with the Lorentz model,
    complementing the earlier points and the Z-scores introduced in our earlier work.\cite{Toraman2025}

    % ... block averages
    \item
    For systems with slow dynamics, the required simulation time in Eq.~\eqref{eq:min-simtime} can easily lead to excessive data storage requirements when data is saved at every time step.
    To reduce the storage burden, we have shown that it is sufficient to keep and analyze block-averaged data.\cite{Toraman2025}
    This is particularly beneficial for high-pressure simulations, where slower system dynamics necessitate much longer trajectories.
    For a block size $B$, every block average stored on disk replaces $B$ individual values it averages over, reducing the data size by a factor of $1/B$.
    Given a time step $h$, one block spans a time interval $Bh$, and according to the Nyquist--Shannon sampling theorem, the highest frequency that can be captured in the averaged data is $1/(2Bh)$.
    To ensure that a Lorentzian at $f=0$ and with width $1/(2\pi \tau_\text{exp})$ can be represented without aliasing artifacts,
    the Nyquist--Shannon frequency of the block-averaged data must be significantly higher than the width of the peak:
    \begin{align*}
        \frac{1}{2Bh} \gg \frac{1}{2\pi \tau_\text{exp}}
    \end{align*}
    Hence, we recommend choosing the maximum block size as:
    \begin{align*}
        B_{\max} \approx \frac{\pi \tau_\text{exp}}{10\,h}
    \end{align*}
    Again, this is not a strict rule but rather a guideline to ensure a good trade-off between storage efficiency and reliable estimates when using STACIE with the Lorentz model.
\end{enumerate}
The Lorentz model implementation in STACIE provides diagnostic guidance on the recommended simulation lengths and block sizes,
enabling a systematic assessment of whether the sampling is sufficient to yield statistically converged estimates.
If the simulation time is insufficient,
the results of the analysis (including the error estimates) are potentially severely biased.
To remedy this, the simulation should be extended further (e.g., by using restart files) until the low-frequency region is sufficiently resolved.

\subsection{Integrated versus exponential correlation time}\label{subsec:tau}

% State the potential for confusion between the two timescales.
The term ``correlation time'' is often used in the context of time-correlated data, and generally refers to the ``integrated correlation time''.
In this work, however, we introduced the ``exponential correlation time'', $\tau_\text{exp}$, in Section~\ref{subsec:lorentz-model}.
In fact, with the Lorentz model, both timescales can be estimated from the same data in a single analysis.
They only coincide in the special case of a purely exponential ACF without additional fast components,
whereas in realistic MD time series, they can differ substantially.
To avoid confusion, this section clarifies the definitions of the two timescales and how they relate to each other.

% Review the integrated correlation time and its relation to the uncertainty in the mean of time-correlated data.
The integrated correlation time appears naturally when quantifying the uncertainty in the sample mean of time-correlated data.
Consider a time series $\{\hat{x}_n\}_{n=0}^{N-1}$ consisting of $N$ time-correlated samples of a scalar observable $x$.
The sample mean and its variance are defined as follows:
\begin{align*}
    \hat{x}_\text{av} &= \frac{1}{N} \sum_{n=0}^{N-1} \hat{x}_n
    \\
    \operatorname{VAR}[\hat{x}_\text{av}] &=
      \frac{1}{N^2} \sum_{n=0}^{N-1} \sum_{m=0}^{N-1}
      \operatorname{COV}[\hat{x}_n \,,\, \hat{x}_m]
\end{align*}
A well-known result from time-series analysis states that, for large $N$, the variance of the sample mean can be approximated as:\cite{Sokal1997, Frenkel2002, Allen2017}
\begin{align*}
    \operatorname{VAR}[\hat{x}_\text{av}]
    = \frac{1}{t_\text{sim}} \int_{-\infty}^\infty c(\Delta_t)\,\mathrm{d} \Delta_t
\end{align*}
This is often rewritten in terms of the integrated correlation time, $\tau_\text{int}$, defined as follows:
\begin{align*}
  \tau_\text{int} = \frac{1}{2} \int_{-\infty}^\infty \frac{c(\Delta_t)}{c(0)}\,\mathrm{d}\Delta_t
\end{align*}
Using this definition and $t_\text{sim} = Nh$, the variance of the sample mean can be expressed in an intuitive form:
\begin{align*}
    \operatorname{VAR}[\hat{x}_\text{av}]
    = \frac{2 \tau_\text{int}}{h} \frac{c(0)}{N}
\end{align*}
The second factor is the naive variance estimate, neglecting time correlations,
and the first factor is called the ``statistical inefficiency'',\cite{Allen2017} which accounts for time correlations.

% Derive the integrated correlation time of the Lorentz model.
To clarify how $\tau_\text{int}$ differs from $\tau_\text{exp}$,
we derive a relation between the two, starting from the Lorentz model.
In the time domain, the corresponding ACF is given by the inverse Fourier transform of $I^\text{lorentz}(f)$:
\begin{align*}
  c(\Delta_t) =
  C_0 \, \delta(\Delta_t)
  +
  C_1 \, \exp\left(-\frac{|\Delta_t|}{\tau_\text{exp}}\right)
\end{align*}
The Dirac delta distribution, $\delta(\Delta_t)$, is a model for the fast dynamics occurring on timescales much shorter than $\tau_\text{exp}$.
In reality, the fast dynamics are not instantaneous and the ACF has a more complex shape at short time lags.
For discrete time series, the fastest oscillations may be limited by the Nyquist frequency.
To show the effect of the fast oscillations on $\tau_\text{int}$, we replace the Dirac delta distribution with a finite sharp peak centered at $\Delta_t=0$:
\begin{align*}
  c(\Delta_t) &=
    S_0 \, f\left(\frac{\Delta_t}{\tau_\text{fast}}\right)
    +
    C_1 \, \exp\left(-\frac{|\Delta_t|}{\tau_\text{exp}}\right)
\end{align*}
where $f$ is a fixed shape function with $f(0) = 1$ and $\int_{-\infty}^{\infty} f(u) \, \mathrm{d} u = 2$.
One can always renormalize $f$, $S_0$ and $\tau_\text{fast}$ to satisfy these conditions.
This model represents a sharp peak if $f(u)$ decays rapidly for $|u| > 1$
and the parameter $\tau_\text{fast}$, which controls the width of the peak, is much shorter than $\tau_\text{exp}$.
Formally, one recovers the original Lorentz model term $C_0 \, \delta(\Delta_t)$ by identifying $C_0 = 2 S_0 \tau_\text{fast}$ and considering the limit $\tau_\text{fast} \to 0$ with $C_0$ held fixed.
With this more realistic ACF, the integrated correlation time can be written as a weighted average of two timescales:
\begin{align*}
  \tau_\text{int}
  =
  \frac{S_0 \tau_\text{fast} + C_1 \tau_\text{exp}}{S_0 + C_1}
\end{align*}
If the ACF has no fast component and is purely exponential ($S_0 = 0$),
the integrated and exponential correlation times coincide, $\tau_\text{int} = \tau_\text{exp}$.
However, in the presence of some fast dynamics, the exponential correlation time always corresponds to the slowest mode and the integrated correlation time becomes a weighted average of multiple timescales present in the ACF.
Note that, because STACIE only fits a model to the spectrum at the lowest frequencies, it can constrain only the effective white-noise background level, $C_0 = 2 S_0 \tau_\text{fast}$.
Neither the detailed shape function $f$ nor the individual parameters $S_0$ and $\tau_\text{fast}$ are identifiable separately from this low-frequency fit.

% Discuss the implications of the derivation.
In summary, $\tau_\text{int}$ and $\tau_\text{exp}$ serve different practical purposes and cannot generally be interchanged.
The two timescales become comparable only in the limit $S_0 \ll C_1$, where this single decay mode determines both the long-time decay and the statistical inefficiency.
In viscosity calculations, deviatoric pressure fluctuations typically exhibit fast oscillations caused by atomic vibrations superimposed on a slow mode.
In such cases, the time series may have a small $\tau_\text{int}$ (hence relatively small uncertainty in time-averaged quantities for a given $t_\text{sim}$) while still exhibiting a large $\tau_\text{exp}$, meaning that long simulations are required to converge the predicted transport properties.

\section{EMD simulations for shear viscosity calculations} \label{sec:emd}
% Overview
This section first outlines the technical details of the EMD simulations from which shear viscosity estimates were derived.
This is followed by a summary of how to transform the microscopic pressure tensor elements, in case of an isotropic liquid, into five statistically independent deviatoric components from which the shear viscosity can be estimated.

\subsection{Simulation details}
\label{subsec:simulation-details}

% Overview of the molecule and the EMD dataset
We estimated the shear viscosity of \textit{2,2,4-trimethylhexane} (C$_9$H$_{20}$), a short, branched hydrocarbon that was extensively studied in the context of the 10$^\text{th}$ IFPSC.\cite{10thIFPSC}
Although this molecule is small compared to base oil lubricant molecules, it does exhibit the piezoviscosity of a realistic lubricant.
It was selected for the 10$^\text{th}$ IFPSC\cite{10thIFPSC} because its small size facilitates convergence in MD simulations and allows for accurate reference shear viscosity measurements to validate the MD predictions.
The availability of experimental data and prior computational studies enables validation and benchmarking of our results.\cite{Kondratyuk2019, Cunha2019, Messerly2019, Gong2019, Zheng2019, Bair2019Pressure}
We analyzed trajectories from an EMD dataset\cite{EMDdata2023} computed using LAMMPS\cite{Thompson2022} with a simulation protocol detailed in our previous work.\cite{Toraman2023}
For this study, we expanded the dataset with simulations at a pressure of \SI{1}{\giga\pascal} using the same protocol,
and made it available in a more reusable format.\cite{EMDdata2025}

% Details of the EMD simulations
Fifty independent NVT EMD simulations were performed for each investigated pressure, with increasing simulation times at higher pressures.
All simulations were conducted at a temperature of \SI{293}{\kelvin} using the Nosé--Hoover thermostat with a relaxation time of \SI{0.5}{\pico\second}.\cite{Fanourgakis2012, Basconi2013, Heinz2026, Nose1984, Hoover1985}
The density of each simulation cell was determined from prior NPT simulations at the target pressure and temperature.\cite{Maginn2020}
At ambient conditions, the shortest simulation time was \SI{2}{\nano\second}, while for high-pressure conditions,
simulations were extended (up to \SI{500}{\nano\second} at \SI{1}{\giga\pascal}).
The simulation duration at \SI{1}{\giga\pascal} was determined by available computational resources.
Each production run at \SI{1}{\giga\pascal} was performed using a single node with a 64-core AMD Epyc 7H12 CPU, and had a wall time of approximately 30 days.
The total CPU hours for the 50 trajectories were approximately 2,300,000 (50 trajectories $\times$ 1 node $\times$ 64 processors $\times$ 30 days $\times$ 24 hours).
(Due to a technical issue, two trajectories failed to complete, yielding 48 independent trajectories at \SI{1}{\giga\pascal}.)

% Dismiss other sources of errors
Since the primary focus of this work is on high-pressure conditions and demonstrating the applicability of STACIE to systems with slow dynamics,
systematic uncertainties and bias due to force field selection or finite-size effects are not accounted for here.\cite{Cunha2019, Gong2019, Messerly2019, Zheng2019}

% Block averaging of pressure tensor elements
To reduce storage demands, particularly at higher pressures, we used the ``fix ave/time'' command of LAMMPS.
In this way, the symmetric microscopic pressure tensor components were block averaged over a chosen block size, $B=2000$,
as explained in Section~\ref{subsec:input-prep}.
Because the MD time step is $h=\SI{0.5}{\femto\second}$, this corresponds to a new average every picosecond.
For each external pressure, the time-dependent block-averaged pressure tensors from 50 independent trajectories were combined,
which served as the inputs for STACIE to estimate shear viscosity and mean pressure, along with their associated uncertainties.

\subsection{Five uncorrelated deviatoric pressure components of an isotropic liquid}\label{subsec:five-uncorrelated}

% Review current state of the art and why we need something new.
Shear viscosity is typically estimated from the off-diagonal elements of the symmetric pressure tensor, $\hat{\mathbf{P}}^{\text{s}}$, where
\begin{align*}
  \begin{aligned}
    \hat{P}^\text{s}_{xy} &= (\hat{P}_{xy} + \hat{P}_{yx})/2
    \\
    \hat{P}^\text{s}_{xz} &= (\hat{P}_{zx} + \hat{P}_{xz})/2
    \\
    \hat{P}^\text{s}_{yz} &= (\hat{P}_{yz} + \hat{P}_{zy})/2
  \end{aligned}
\end{align*}
and the final viscosity is expressed as an average over the three estimates.
LAMMPS only prints out the symmetric components, as these are relevant for post-processing.\cite{Evans2008, Thompson2022}
It is well established that the diagonal elements also contain relevant information for the shear viscosity,\cite{Mondello1997, Borodin2009, Liu2012, Aquing2012, MercierFranco2023} and using them is recommended by the best practices guide by Maginn and co-workers.\cite{Maginn2020}

% Explain Daivis-Evans
For isotropic fluids, Daivis and Evans showed that the shear viscosity can be estimated using the full orthogonal symmetric (traceless symmetric) pressure tensor, $\hat{\mathbf{P}}^\text{os}$:\cite{Daivis1994}
\begin{align}
  \eta = \frac{V}{10 k_\text{B} T} \frac{1}{2}\int_{-\infty}^\infty \left\langle
    \hat{\mathbf{P}}^\text{os}(t_0) : \hat{\mathbf{P}}^\text{os}(t_0 + \Delta_t)
  \right\rangle \mathrm{d}\Delta_t
  \label{eq:daivis-evans}
\end{align}
with
\begin{align*}
  \hat{\mathbf{P}}^\text{os} = \hat{\mathbf{P}}^\text{s} - \frac{1}{3} \text{Tr}(\hat{\mathbf{P}}) \mathbf{I}
\end{align*}
Although the method of Daivis and Evans enhances statistical accuracy, removing the trace introduces correlations among the diagonal pressure tensor elements.
This conflicts with STACIE's requirement of statistically independent inputs,
on which it relies to perform uncertainty quantification.
Consequently, the approach by Daivis and Evans cannot be combined with STACIE directly.

% Review of literature on transformations of the pressure tensor
Another popular approach for isotropic fluids is to include linear combinations of diagonal elements:
\begin{align*}
  \begin{aligned}
    \hat{\Pi}_{xx} &= (\hat{P}_{yy} - \hat{P}_{zz})/2
    \\
    \hat{\Pi}_{yy} &= (\hat{P}_{zz} - \hat{P}_{xx})/2
    \\
    \hat{\Pi}_{zz} &= (\hat{P}_{xx} - \hat{P}_{yy})/2
  \end{aligned}
\end{align*}
These appear as off-diagonal elements after a \SI{45}{\degree} rotation of the coordinate system about the $x$, $y$, or $z$ axis, respectively.
Holian and Evans proposed including all three,\cite{Holian1983} despite their linear dependence, which implies that all information is contained in only two of them.
To accommodate this concern, Alfé and Gillan suggested using only two of the linear combinations.\cite{Alfe1998}
Many authors refer to the resulting set, $\{\hat{P}^\text{s}_{xx}, \hat{P}^\text{s}_{yy}, \hat{P}^\text{s}_{zz}, \hat{\Pi}_{xx}, \hat{\Pi}_{yy}\}$ or trivial variations thereof, as ``the five independent components of the traceless virial stress tensor''.\cite{Padding2001, Fanourgakis2012, Mouas2012, Raabe2012, Pozzo2013, Guillaud2017a, Damone2019}
This is potentially confusing because $\hat{\Pi}_{xx}$ and $\hat{\Pi}_{yy}$ are \textit{linearly} independent but not \textit{statistically} independent.
Assuming that the original diagonal elements are independent and identically distributed, the Pearson correlation coefficient between all pairs in $\{\hat{\Pi}_{xx}, \hat{\Pi}_{yy}, \hat{\Pi}_{zz}\}$ is $-1/2$.
Hence, this set of five (or six) components cannot be used as inputs for STACIE either.

\begin{figure}
  \centering
  \includegraphics{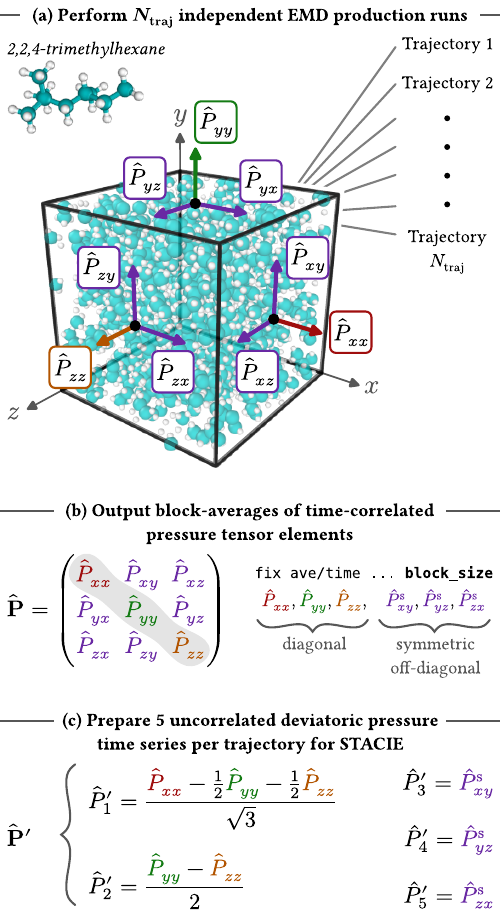}
  \caption{
    Preparation of five uncorrelated deviatoric pressure components per trajectory as input for STACIE's shear viscosity calculation.
    (a) Illustration of pressure tensor elements for a cubic simulation box containing 100 \textit{2,2,4-trimethylhexane} molecules.
    (b) Block averages of pressure tensor elements printed using the ``fix ave/time'' command in LAMMPS.\cite{Thompson2022}
    (c) Transformation of the pressure tensor into the five uncorrelated components required as input for STACIE, resulting in $M=5 N_\text{traj}$ input time series in total.
  }
  \label{fig:stress-tensor-components}
\end{figure}

% Introduce a new set of five uncorrelated deviatoric pressure components
In Section~S2 of the Supporting Information, we show that the expression of Daivis and Evans can be rewritten as an average over five statistically independent estimates of the viscosity, each obtained from a different linear combination of the pressure tensor elements.
To avoid confusion, we refer to them as ``the five \textit{uncorrelated} deviatoric pressure components'':
\begin{align}
  \begin{aligned}
    \hat{P}'_1 &= \frac{\hat{P}_{xx} - (\hat{P}_{yy} + \hat{P}_{zz}) / 2}{\sqrt{3}}
    \\
    \hat{P}'_2 &= \frac{\hat{P}_{yy} - \hat{P}_{zz}}{2}
    \\
    \hat{P}'_3 &= \hat{P}^\text{s}_{yz}
    \\
    \hat{P}'_4 &= \hat{P}^\text{s}_{zx}
    \\
    \hat{P}'_5 &= \hat{P}^\text{s}_{xy}
  \end{aligned}
  \label{eq:deviatoric-components}
\end{align}
Figure~\ref{fig:stress-tensor-components} illustrates the pressure tensor elements on a cubic simulation box of \textit{2,2,4-trimethylhexane}
and shows how to prepare them as an input for STACIE.
In their best-practices guide, Maginn \textit{et al.} stated that the precise statistical advantage of the method of Daivis and Evans was not clear.\cite{Maginn2020}
By rewriting Eq.~\eqref{eq:daivis-evans} in terms of these five components, this advantage becomes evident:
it reduces the uncertainty of the viscosity estimate by a factor of $\sqrt{3/5}$, which corresponds to a reduction of about 23\%, compared to using only the three off-diagonal elements.

% Literature survey on the use of pressure tensor components
The literature on the use of pressure tensor components in EMD-based shear viscosity calculations is fragmented, and one does not always clearly specify which components were used.
We conducted an extensive literature survey to assess the current practices and to identify prior publications that may already have introduced the five uncorrelated deviatoric pressure components.
A full account of this survey is provided in Section~S3 of the Supporting Information.
We identified 334 papers that clearly described the choice of pressure tensor components used in the viscosity calculation.
This survey revealed that the most common approach by far (50\% of the papers surveyed) is to use only the three off-diagonal elements and discard the diagonal elements.
Most of the remaining papers are almost equally divided between those employing the method of Daivis and Evans and those that use linear combinations of diagonal elements (24\% each).
Only five papers were found to contain elements similar to the five components proposed here.\cite{Voadlo2000, Dai2018, Dai2023, Wang2023, Zhang2025}
Thus, while the five deviatoric components defined in Eq.~\eqref{eq:deviatoric-components} are not entirely unprecedented, they have not previously been formulated explicitly and analyzed in terms of their statistical benefits.
More importantly, their adoption is practically non-existent, which is a missed opportunity.
We hope that our analysis clarifies how one can easily improve the accuracy of any type of EMD-based shear viscosity calculation, by using the five uncorrelated deviatoric pressure components,
or the expression of Daivis and Evans.

\section{Shear viscosity of \textit{2,2,4-trimethylhexane}} \label{sec:results}
% Overview paragraph
We first apply the Lorentz model in STACIE to analyze the viscosity of \textit{2,2,4-trimethylhexane} at ambient conditions, and illustrate some of the intermediate results provided by STACIE.
Subsequently, we extend our analysis to higher pressures, up to \SI{1}{\giga\pascal}, and compare our findings with experimental data as well as results from other MD studies in the literature.

In principle, a direct comparison to the experimental viscosity is only possible when the simulations are performed at exactly the same pressure as the experiments.
However, the average pressure in NVT simulations can slightly deviate from the target value, and due to the strong piezoviscous effect, even small pressure differences can result in significant changes in viscosity.
To address this issue, all comparisons are made using the hybrid McEwen--Paluch model, $\eta_\text{hyb}(P)$, fitted by Bair\cite{Bair2019Pressure} to the experimental data, and evaluated at the average pressure of our production runs.
To clarify the significance of any discrepancies, we also constructed a simple model for the standard uncertainty of the experimental viscosity as a function of pressure.
An analysis of the McEwen--Paluch model and the experimental uncertainties is given in Section~S4 of the Supporting Information.

\subsection{Ambient conditions}\label{sec:ambient}

\begin{figure}
  \includegraphics{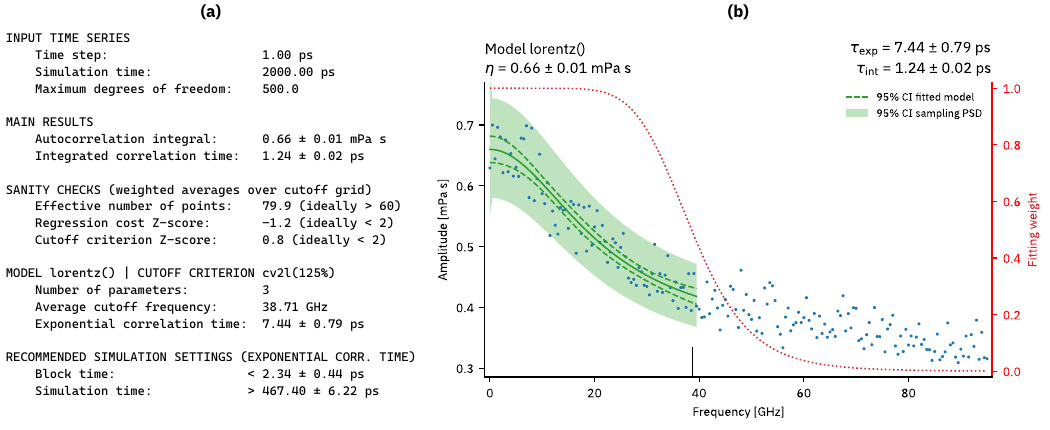}
  \caption{
    STACIE viscosity analysis of \textit{2,2,4-trimethylhexane} at ambient conditions (\SI{0.1}{\mega\pascal} and \SI{293}{\kelvin}),
    with 50 trajectories, each \SI{2}{\nano\second} long.
    (a) The final part of STACIE's screen output showing the shear viscosity estimate and recommended simulation settings.
    Note that the ``time step'' is not the MD integration time step, but the block size in units of time, as this is the input time series given to STACIE.
    (b) PSD averaged over the five uncorrelated deviatoric components and 50 MD trajectories, fitted with the Lorentz model.
    A detailed description of all elements of the spectrum plot can be found in Section~S5 of the Supporting Information.
  }\label{fig:stacie-ambient}
\end{figure}

% Illustration of STACIE's outputs at ambient conditions.
Figure~\ref{fig:stacie-ambient} presents STACIE's output for the shear viscosity at ambient conditions (\SI{0.1}{\mega\pascal} and \SI{293}{\kelvin}).
The screen output shows the main results and the recommended simulation settings.
The graph displays the average spectrum of the uncorrelated deviatoric components and the Lorentz model fitted to these data.
As can be seen from the graph, the low-frequency part showed no indication of the characteristic $a + b \sqrt{f}$ dependence associated with a $\Delta_t^{-3/2}$ power-law decay of the ACF.\cite{Holian1983}
The estimated shear viscosity using \SI{2}{\nano\second} trajectories was $\eta = \SI{0.66 \pm 0.01}{\milli\pascal\second}$,
closely matching the experimental value of $\eta_\text{hyb} = \SI{0.65 \pm 0.01}{\milli\pascal\second}$.\cite{Bair2019Pressure}

% Comment on sufficiency of the simulation time.
As discussed in Section~\ref{subsec:input-prep}, the exponential correlation time, $\tau_\text{exp}$, is the crucial parameter to confirm that the simulation time is sufficient.
Figure~\ref{fig:stacie-ambient}(b) shows that $\tau_\text{exp}$ is estimated to be $\SI{7.44}{\pico\second}$.
Applying our heuristic guideline in Eq.~\eqref{eq:min-simtime}, the minimal recommended simulation time is $20 \times \pi \times \tau_\text{exp} \approx $ \SI{0.47}{\nano\second},
confirming that our simulation time of \SI{2}{\nano\second} is sufficient to capture the slowest decay mode in the system for this condition.
Additionally, the recommended maximal block size is \SI{2.34}{\pico\second}, indicating that our selected block size of \SI{1}{\pico\second} is small enough to obtain a low-frequency spectrum without aliasing artifacts.

% Comment on the integrated correlation time.
Note that the integrated correlation time, $\tau_\text{int}$, is estimated to be $\SI{1.22}{\pico\second}$, roughly six times smaller than $\tau_\text{exp}$.
This reveals that the two time constants can indeed differ significantly,
meaning that the slow dynamics of interest is superimposed with faster decay modes.
Some caution is needed when interpreting $\tau_\text{int}$ as a physical timescale of the system.
It is unavoidably affected by the block averaging procedure, which filters out high-frequency modes that would otherwise reduce the integrated correlation time.
Hence, $\tau_\text{int}$ cannot go below the block time, while $\tau_\text{exp}$ remains unaffected by this choice.

\subsection{Pressure dependence of the viscosity up to 1 GPa}\label{sec:high-pressure}

% Provide some context on the high-pressure viscosity of lubricants.
A key property of liquid lubricants is the pressure-viscosity relationship,
which describes the substantial increase in the viscosity of liquid lubricants with increasing pressure.
This trend is particularly pronounced under elastohydrodynamic lubrication (EHL) conditions,
where pressure-induced densification and reduced molecular mobility significantly affect the lubricant film thickness in highly loaded contacts.
This section investigates this piezoviscous effect for \textit{2,2,4-trimethylhexane} at pressures up to \SI{1}{\giga\pascal}.
Our simulation results are compared against experimental data from the 10$^\text{th}$ IFPSC,
alongside computational findings from contributing researchers.\cite{Bair2019Pressure, Zheng2019, Gong2019, Kondratyuk2019, Messerly2019, Cunha2019}
These comparative studies employed various methodologies, including both EMD and NEMD, and different force fields.
As previously stated, we used the COMPASS class II force field\cite{Sun1998} in our simulations.\cite{Toraman2023}

\begin{figure}
  \includegraphics{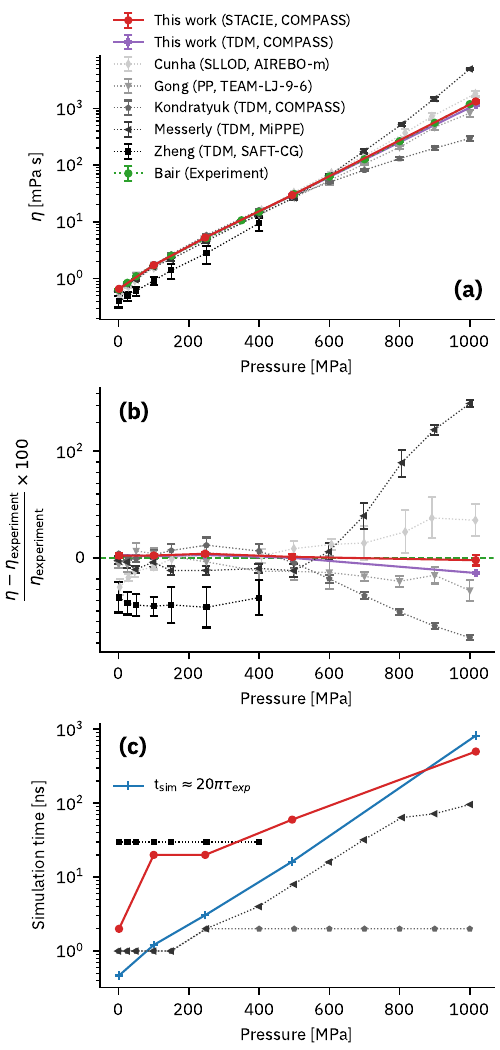}
  \caption{
    Comparison of shear viscosity results with experimental data\cite{Bair2019Pressure} and entries from the 10$^\text{th}$ IFPSC.\cite{Bair2019Pressure, Zheng2019, Gong2019, Kondratyuk2019, Messerly2019, Cunha2019}
    (a) Shear viscosity as a function of pressure up to \SI{1}{\giga\pascal}.
    (b) Relative error in viscosity versus pressure. The dashed green line indicates the experimental value as the zero-reference.\cite{Bair2019Pressure}
    A ``symlog'' scale with a linear threshold of 100 is used on the y-axis to emphasize discrepancies.
    (c) EMD simulation times as a function of pressure, with the solid blue line representing the minimum required simulation time based on Eq.~\eqref{eq:min-simtime}.
    The legend in (a) applies to all subplots, with consistent gray shades used for each IFPSC entry.
    Lines connecting data points are guides to the eye.
    Results obtained in this work are plotted in red (STACIE) and purple (TDM).
    Experimental data are plotted in green.
    SLLOD\cite{Evans1984} and PP\cite{Hess2002} are NEMD methods.
    COMPASS,\cite{Sun1998} AIREBO-m\cite{OConnor2015}, TEAM-LJ-9-6,\cite{Gong2019} MiPPE,\cite{Messerly2019} and SAFT-CG\cite{Muller2014}
    are the force fields used in the MD simulations.
  }\label{fig:pressure-viscosity}
\end{figure}

% Describe the figures and tables showing the results.
Figure~\ref{fig:pressure-viscosity}(a) presents the viscosity as a function of pressure, comparing STACIE's results with those from other studies.
In addition, we have analyzed our MD trajectories using an implementation of the TDM from our previous work with its default settings.\cite{Zhang2015,Toraman2023}
When available, we used the average pressure obtained from the production runs, $\bar{P}_\text{MD}$ (this work, Cunha \textit{et al.},\cite{Cunha2019} Messerly \textit{et al.}\cite{Messerly2019}).
For the other studies,\cite{Zheng2019, Gong2019, Kondratyuk2019} we used the desired pressure from the NPT simulations, $P_D$.
To display the discrepancies better, Figure~\ref{fig:pressure-viscosity}(b) shows the relative error in viscosity compared to the hybrid McEwen--Paluch model.
The error bars in Fig.~\ref{fig:pressure-viscosity}(b) account for propagated uncertainties from simulations and experimental measurements.
Viscosity values obtained with STACIE at different pressures are compared with experimental values and the hybrid McEwen--Paluch model in Table~\ref{tab:results}.
Detailed plots with intermediate results of STACIE at each pressure can be found in Section~S5 of the Supporting Information.

% Comment on pressure uncertainty
We also applied STACIE to estimate the sampling uncertainty in the average pressure,
which was found to be at least five orders of magnitude smaller than the average.
The discrepancy between the average and desired pressures in Table~\ref{tab:results}
stems entirely from a small error in the density,
which was estimated from NPT simulations preceding the NVT production runs.

% Summarize the main findings of STACIE and comparison to other studies.
The results shown in Figure~\ref{fig:pressure-viscosity} indicate that many IFPSC challenge participants accurately estimated shear viscosity up to \SI{500}{\mega\pascal},
except for Zheng \textit{et al.},\cite{Zheng2019}
whose coarse-grained interaction model likely introduces systematic errors.
In contrast, results obtained with STACIE show an excellent agreement with experimental data across the entire pressure range.
The relative error is below \SI{4}{\percent} at all pressures considered and is comparable to the predicted relative uncertainty.
Our results obtained with TDM closely align with STACIE's estimates, except at \SI{1}{\giga\pascal}, where the TDM clearly underestimates the viscosity.

\begin{table}
  \caption{
    Shear viscosity ($\eta$) and exponential correlation time ($\tau_\text{exp}$) for \textit{2,2,4-trimethylhexane} at various pressures, estimated using STACIE.
    $P_D$ is the desired pressure in the NPT equilibration runs, while $\bar{P}_\text{MD}$ is the average pressure in the NVT production runs performed at density $\rho$.
    Experimental viscosities, $\eta_\text{experiment}$, at pressures $P_\text{D}$, are taken from Bair \textit{et al.}\cite{Bair2019Pressure}
    The hybrid McEwen--Paluch model, $\eta_\text{hyb}$, fitted to the experimental data by Bair \textit{et al.}\cite{Bair2019Pressure} is evaluated at $\bar{P}_\text{MD}$ for a more precise comparison with our simulation results.
  }
  \resizebox{\textwidth}{!}{
    
\centering
\begin{tabular}{
    S[table-format=4.1]
    S[table-format=4.3(6)]
    S[table-format=4.2]
    S[table-format=3.0]
    S[table-format=4.3(5)]
    S[table-format=4.3(5)]
    S[table-format=5.3(7)]
}
    \toprule
    \multicolumn{1}{c}{$P_{D}$ [\si{\mega\pascal}]} &
    \multicolumn{1}{c}{$\eta_\text{experiment}$ [\si{\milli\pascal\second}]} &
    \multicolumn{1}{c}{$\bar{P}_\text{MD}$ [\si{\mega\pascal}]} &
    \multicolumn{1}{c}{$\rho$ [\si{\kilogram\per\meter^3}]} &
    \multicolumn{1}{c}{$\eta_\text{hyb}(\bar{P}_\text{MD})$ [\si{\milli\pascal\second}]} &
    \multicolumn{1}{c}{$\eta$ [\si{\milli\pascal\second}]} &
    \multicolumn{1}{c}{$\tau_\text{exp}$ [\si{\pico\second}]}
    \\
    \midrule

    0.1 &
    0.64(0.01) &
    1.14 &
    722 &
    0.65(0.01) &
    0.66(0.01) &
    7.44(0.79)
    \\

    100.0 &
    1.71(0.02) &
    99.89 &
    784 &
    1.69(0.02) &
    1.73(0.02) &
    19.24(1.48)
    \\

    250.0 &
    5.13(0.06) &
    246.50 &
    834 &
    5.11(0.06) &
    5.30(0.07) &
    48.68(4.18)
    \\

    500.0 &
    30.90(0.49) &
    494.17 &
    888 &
    29.37(0.43) &
    29.62(0.52) &
    258.44(21.96)
    \\

    1000.0 &
    1187.00(26.71) &
    1017.07 &
    959 &
    1357.06(28.76) &
    1325.97(61.00) &
    13071.65(2754.74)
    \\

    \bottomrule
\end{tabular}

  }
  \label{tab:results}
\end{table}

% Hypothesize on the cause of discrepancies in other studies: too short simulation times.
Insufficient simulation time is one potential cause of the discrepancies between the simulated and experimental viscosities reported in other studies.
As pressure increases, the exponential time constant, $\tau_\text{exp}$, also increases, requiring much longer simulations to accurately capture the slowest decay mode.
This problem is evident in Figure~\ref{fig:pressure-viscosity}(c), where many entries fall short of the required simulation time, especially at higher pressures.
Among the challenge entries, Messerly \textit{et al.}\cite{Messerly2019} used the longest simulation time at \SI{1}{\giga\pascal} (\SI{96}{\nano\second}),
but overestimated the viscosity by nearly \SI{300}{\percent}.
As noted by the authors, this discrepancy is likely due to the Mie 16-6 potential used in their work, which is overly repulsive at short intermolecular distances.\cite{Messerly2019}
Zheng \textit{et al.}\cite{Zheng2019} performed \SI{40}{\nano\second} simulations at each pressure,
but their coarse-grained model resulted in solidification above \SI{500}{\mega\pascal}.

% Discuss the results at 1 GPa in detail.
At \SI{1}{\giga\pascal}, the simulation time employed in this work (\SI{500}{\nano\second}) was substantially longer than those reported in previous studies.
As expected, the system exhibited slow dynamics, with an exponential correlation time of approximately \SI{13.07}{\nano\second},
implying a recommended simulation duration of around \SI{830}{\nano\second}.
Nevertheless, STACIE predicted a viscosity of $\eta = \SI{1326 \pm 61}{\milli\pascal\second}$ at \SI{1017}{\mega\pascal},
which is in excellent agreement with the hybrid model fitted to the experimental data at the same pressure, $\eta_\text{hyb} = \SI{1357 \pm 39}{\milli\pascal\second}$.
Further extending the simulation time could improve convergence of our results as it would increase the resolution of the low-frequency region of the spectrum.
Unfortunately, this was not feasible due to computational resource limitations.
For comparison, Kondratyuk \textit{et al.}\cite{Kondratyuk2019} used the same force field but conducted much shorter production runs (\SI{2}{\nano\second}) at pressures above \SI{250}{\mega\pascal},
resulting in viscosity underestimations of roughly 75\% at \SI{1}{\giga\pascal}.
The authors attributed this discrepancy to the limitations of the force field.
Our findings, however, indicate that insufficient simulation time is likely the dominant factor affecting the accuracy under such conditions,
rather than the deficiencies in the force field itself.

\begin{figure}
  \includegraphics{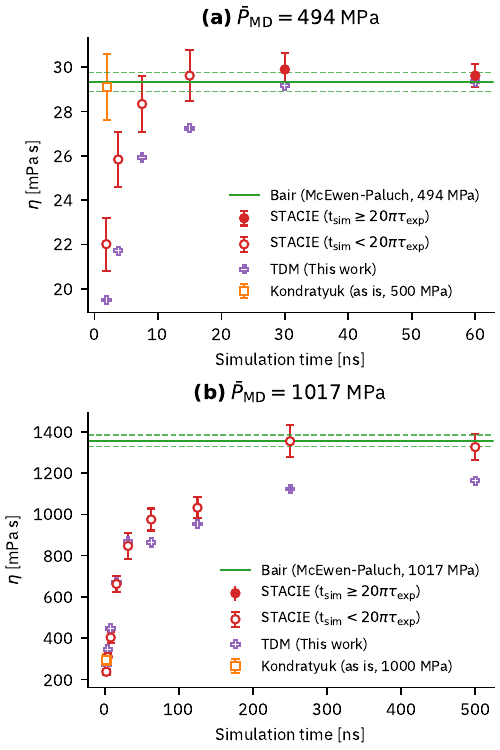}
  \caption{
    Shear viscosity with its standard uncertainty estimated from EMD trajectories as a function of simulation time.
    (a) $\bar{P}_\text{MD} = \SI{494}{\mega\pascal}$ with \SI{60}{\nano\second}-long trajectories.
    (b) $\bar{P}_\text{MD} = \SI{1017}{\mega\pascal}$ with \SI{500}{\nano\second}-long trajectories.
    The green solid and dashed lines represent the experimental value and its standard uncertainty, respectively,
    evaluated at the average pressure of our production runs using the hybrid McEwen--Paluch model.\cite{Bair2019Pressure}
    Red points show the viscosity estimates obtained with STACIE for truncated trajectories, with hollow symbols indicating estimates that do not meet the minimum simulation time recommended by STACIE.
    Purple plusses correspond to viscosity estimates from this work using TDM.
    Orange squares correspond to viscosity estimates of Kondratyuk \textit{et al.}.\cite{Kondratyuk2019}
  }\label{fig:convergence}
\end{figure}

% Convergence tests by truncating the trajectories.
The importance of simulation time is further supported by convergence tests in which we estimated the viscosity from truncated trajectories.
Starting from the full trajectory, we discarded the second half, analyzed the first half, and then repeated this procedure iteratively until the retained part was shorter than \SI{2}{\nano\second}.
Figure~\ref{fig:convergence}(a) shows the estimated shear viscosity at \SI{494}{\mega\pascal} as a function of simulation time,
using truncated sequences with durations $\{1.875, 3.75, 7.5, 15, 30, 60\}$~\SI{}{\nano\second}.
Figure~\ref{fig:convergence}(b) presents the same analysis for \SI{1017}{\mega\pascal},
with simulation times of $\{1.95, 3.9, 7.8, 15.6, 31.25, 62.5, 125, 250, 500\}$~\SI{}{\nano\second}.
Hollow symbols indicate results for which the minimum simulation time recommended by STACIE was not met.
In such cases, the viscosity estimate and its uncertainty are not expected to be reliable.
Plots with STACIE's intermediate results for the truncated trajectories are given in Section~S6 and S7 of the Supporting Information.
For \SI{494}{\mega\pascal}, the viscosity estimate converges towards the experimental value after about \SI{15}{\nano\second},
whereas at \SI{1017}{\mega\pascal}, it only converges to the experimental value after about \SI{125}{\nano\second}.

% Discuss the TDM results in the context of the convergence tests.
Figure~\ref{fig:convergence} also includes TDM viscosity estimates, reported by Kondratyuk \textit{et al.}\cite{Kondratyuk2019} (orange squares) and obtained in this work (purple plusses).
Figure~\ref{fig:convergence}(a) shows that TDM results may vary significantly for the same pressure and simulation time.
We applied TDM using its default settings,\cite{Zhang2015,Toraman2023} whereas Kondratyuk \textit{et al.}\cite{Kondratyuk2019} applied TDM with increased cutoffs during the regression of the double-exponential model.
This sensitivity has been noted previously\cite{Messerly2019,Toraman2023} and limits the relevance of direct comparisons with TDM results, since the method can be tuned to better match experimental data through adjustment of its hyperparameters.
Figure~\ref{fig:convergence}(b) shows that the default TDM settings are not universally applicable and produce systematic errors at high pressures, even with extended simulation times.
These observations suggest that TDM generally requires hyperparameter tuning and visual inspection of the fitted models to obtain reliable viscosity estimates.

% Describe extrapolation method to even higher pressures.
Finally, assuming that the shear viscosity continues to increase exponentially with pressure
and that the lubricant remains in a liquid state
(which is not necessarily guaranteed at such high pressures),
we provide a rough estimate of the simulation time required to obtain converged results
at \SI{1.5}{\giga\pascal} and \SI{2}{\giga\pascal}.
This extrapolation is not intended as a definitive prediction of viscosity,
but rather as an illustration of the computational challenges of lubricant EMD simulations under such extreme conditions.

\begin{figure}
  \centering
  \includegraphics{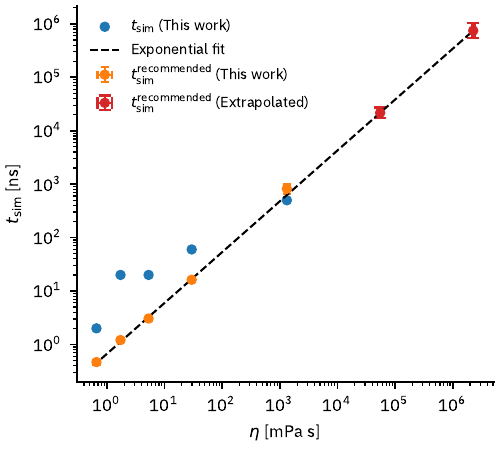}
  \caption{
    Simulation times as a function of viscosity.
    Blue points show the simulation times used in this study to compute viscosities up to \SI{1}{\giga\pascal}.
    STACIE's recommended simulation times based on the exponential correlation time, $\tau_\text{exp}$, are shown in orange.
    Red points represent the extrapolated viscosities and corresponding simulation times at \SI{1.5}{\giga\pascal} and \SI{2}{\giga\pascal},
    assuming an exponential viscosity increase with pressure.
    The fitted model is shown as a dashed black line.
  }\label{fig:extrapolation}
\end{figure}

% Show and discuss the extrapolation results.
The results of the extrapolation are shown in Figure~\ref{fig:extrapolation}.
The error bars only represent the sampling uncertainty propagating through the regression
from the viscosity estimates at lower pressures.
They do not account for potential systematic errors in the exponential model.
At \SI{1.5}{\giga\pascal} and \SI{2}{\giga\pascal}, the extrapolated viscosity values reached the order of 10$^5$ and 10$^6$ \SI{}{\milli\pascal\second}, respectively.
Correspondingly, the required simulation times extended into the microsecond and millisecond ranges.
In principle, such extended simulations are possible with advanced supercomputing and GPU acceleration, but they remain extremely computationally demanding.

% Discuss some caveats of the extrapolation.
The results of this extrapolation should be interpreted with caution.
At sufficiently high pressures, liquid lubricants may undergo a transition to a glassy or amorphous solid-like state,
typically occurring at a pressure of about \SI{2}{\giga\pascal}.\cite{Bair2019High}
Under such conditions, molecular mobility is severely restricted,
which can lead to a dramatic increase in apparent viscosity.\cite{Bair2016}
Furthermore, the assumptions underlying standard force fields parameterized for the liquid state may no longer hold at extreme conditions.\cite{Ponder2010, Jorgensen1996}

% Summary of the results and discussion.
In summary, our findings demonstrate that EMD can produce reliable high-pressure viscosity estimates, albeit at a significant computational cost.
By applying STACIE, we achieved, to the best of our knowledge, the most accurate EMD-based shear viscosities reported for \textit{2,2,4-trimethylhexane}.
Throughout this work, we highlighted several key considerations for using EMD to calculate liquid viscosity at very high pressures:
the need for sufficiently long trajectories to capture the slowest decay mode, which requires accurate estimation of the exponential correlation time, $\tau_\text{exp}$,
and the reduction of storage requirements through block averages.
These insights and best practices provide a framework for investigating lubricant rheology under high pressures,
optimizing the trade-off between computational efficiency and predictive accuracy.

\subsection{Statistical independence of the five deviatoric components}\label{subsec:validation-five}

% Overview of the validation of the lack of correlation, some background
In Section~\ref{subsec:five-uncorrelated}, we proposed a set of five uncorrelated deviatoric pressure components to estimate the viscosity.
For completeness, we also validated their statistical independence numerically, using all the trajectory data in this work.
To ensure that this analysis is mainly sensitive to the slowest oscillations in the pressure tensor, we preprocessed the pressure time series by block-averaging them with a block size equal to $\tau_\text{exp}$, resulting in $N_b$ blocks for each component and each trajectory.
We then computed the Pearson correlation coefficient $r$ for each pair of components, and for all trajectories.
These sets of correlation coefficients are themselves random variables and were analyzed using the Fisher transformation:\cite{Fisher1915}
for sufficiently large $N_b$, the distribution of $\operatorname{atanh}(r)$ is known to follow a normal distribution with standard deviation $1/\sqrt{N_\text{b}-3}$, irrespective of the expected value of the correlation coefficient.

\begin{figure}
  \includegraphics{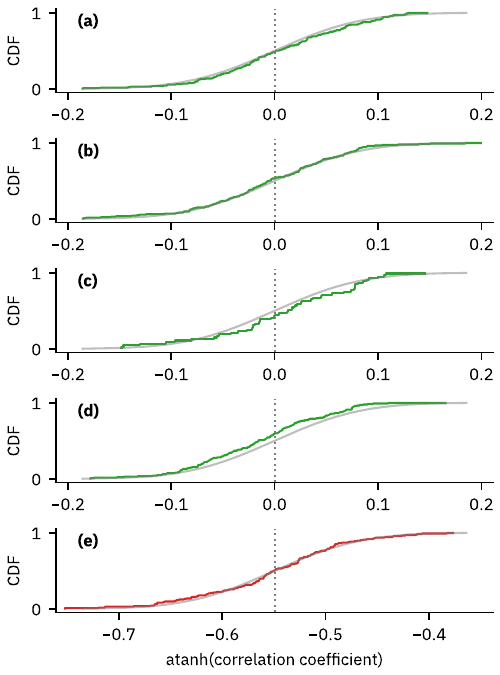}
  \caption{
    Cumulative distributions of the Fisher-transformed Pearson correlation coefficients between pairs of pressure components. Data were grouped by permutational equivalence:
    (a) $\hat{P}'_1(t)$ and one of $\{\hat{P}'_3(t), \hat{P}'_4(t), \hat{P}'_5(t)\}$,
    (b) $\hat{P}'_2(t)$ and one of $\{\hat{P}'_3(t), \hat{P}'_4(t), \hat{P}'_5(t)\}$,
    (c) $\hat{P}'_1(t)$ and $\hat{P}'_2(t)$,
    (d) all pairs in $\{\hat{P}'_3(t), \hat{P}'_4(t), \hat{P}'_5(t)\}$,
    and (e) all pairs in $\{\hat{\Pi}_{xx}(t), \hat{\Pi}_{yy}(t), \hat{\Pi}_{zz}(t)\}$.
    Empirical distributions shown in color, analytical distributions shown in grey.
    The dotted horizontal line is the expected value, zero for panels (a) to (d) and $\operatorname{atanh}(-1/2)$ for panel (e).
  }\label{fig:correlation}
\end{figure}

% Results showing the (absence of) correlation
Panels (a) to (d) of Figure~\ref{fig:correlation} show distributions of the Fisher-transformed correlation coefficients, for different subsets of pressure components,
from the $P=\SI{500}{\mega\pascal}$ simulations.
(Similar plots for other pressures are given in Section~S8 of the Supporting Information.)
The transformed correlation coefficients follow the expected normal distribution with zero mean.
Figure~\ref{fig:correlation}(e) shows the distribution of correlation coefficients between pairs of $\{\hat{\Pi}_{xx}, \hat{\Pi}_{yy}, \hat{\Pi}_{zz}\}$ time series.
These components are clearly correlated, with the expected Pearson correlation coefficient of $-1/2$.

\begin{figure}
  \includegraphics{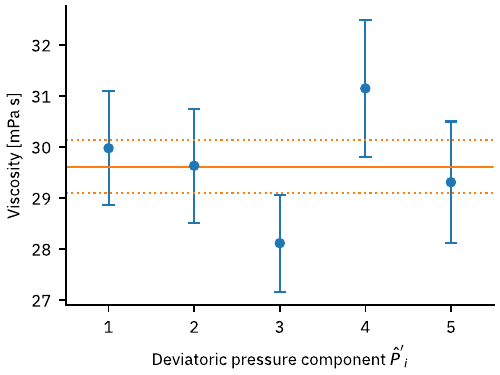}
  \caption{
    Shear viscosity estimates obtained using only one of the five deviatoric pressure components at a time, for $P=\SI{500}{\mega\pascal}$.
    The estimate obtained using the combined components is shown as a solid horizontal line for comparison.
    Error bars and dashed lines represent the standard error.
    In all five cases, all of STACIE's sanity checks passed, confirming that the results and the error estimates are reliable.
  }\label{fig:viscosity-separate}
\end{figure}

% Also show viscosities obtained by using only one of the five contributions.
To further validate the proposed deviatoric pressure components, we also estimated the viscosity using only one component at a time.
The results, again for the $P=\SI{500}{\mega\pascal}$ simulations, are shown in Figure~\ref{fig:viscosity-separate}.
Viscosity estimates obtained using the separate components are consistent with each other and with the estimate obtained from the combined components.
Similar plots for other pressures are given in Section~S8 of the Supporting Information.
For some conditions, notably $P=\SI{1}{\giga\pascal}$, a single component may not provide sufficient information to obtain a reliable viscosity estimate, as indicated by the sanity checks implemented in STACIE.

\section{Conclusions} \label{sec:conclusions}
% Very short summary of the most important findings
We introduced the Lorentz model for estimating transport properties with STACIE.
Using this model, we calculated the shear viscosity of a model molecule, \textit{2,2,4-trimethylhexane} at pressures up to \SI{1}{\giga\pascal} via EMD simulations.
Our results show excellent agreement with experimental data at high pressures, validating the method's accuracy.
Furthermore, a comparison with the 10$^\text{th}$ IFPSC emphasized that high-viscosity liquids require long simulation times.

% Detailed conclusions on the Lorentz model
The applicability of the Lorentz model extends beyond shear viscosity.
For instance, we used it to quantify the uncertainty in the mean pressure in our simulations.
STACIE's documentation details additional use cases, including bulk viscosity and thermal conductivity calculations.
The model is suitable whenever the relevant autocorrelation function exhibits exponential decay.
A key advantage of the Lorentz model is its ability to estimate the exponential correlation time, $\tau_\text{exp}$, which characterizes the system's slowest timescale.
If faster modes are also present, $\tau_\text{exp}$ is greater than the commonly used integrated correlation time, $\tau_\text{int}$.
With $\tau_\text{exp}$, one can derive the minimum simulation time required to obtain reliable transport properties and their associated uncertainties.
Additionally, $\tau_\text{exp}$ can guide the size of block averages for writing out dynamical quantities, thereby optimizing storage efficiency without compromising post-processing quality.

% Conclusions on the five deviatoric components
Because STACIE requires independent time series as inputs,
we analyzed how the Cartesian pressure tensor can be transformed into five uncorrelated deviatoric components.
Two components are derived from the diagonal components of the pressure tensor, while the remaining three are off-diagonal components.
We demonstrated, both theoretically and numerically, that these five components are statistically independent and all equally valid separately for viscosity estimation.
Averaging the viscosity over these five components is equivalent to the method of Daivis and Evans, and is statistically optimal.

% Extrapolation and outlook on future MD studies
Finally, we extrapolated the shear viscosity of \textit{2,2,4-trimethylhexane} to pressures beyond \SI{1}{\giga\pascal} and estimated the required simulation times, offering valuable insights into the computational challenges facing future viscosity calculations.
In future work, we will also apply STACIE to longer molecules that are more representative of industrial lubricants.
Although STACIE has proven robust and reliable, the computational cost of EMD simulations for longer chains and higher pressures will likely be significant.
Nonetheless, post-processing should remain feasible provided that time-correlated data is acquired following the procedures outlined in this work.

% Methodological outlook
Methodologically, it would be beneficial to extend STACIE beyond the maximum a posteriori (MAP) estimate of model parameters and their covariance.
The current implementation of the Lorentz model imposes a penalty to mitigate the influence of models fitted to insufficient spectral data, where poorly resolved parameters may violate MAP assumptions.
A more rigorous analysis of the full posterior distribution, for instance, through Markov Chain Monte Carlo (MCMC) sampling, could improve STACIE's robustness, especially for cases with limited data.
In addition, future versions of STACIE could be extended with additional spectral models, e.g. the Fourier transform of the stretched exponential model, to better describe the slow decay associated with the transient power-law regime of the ACF.

\section*{Data and software availability} \label{sec:availability}
The method used in this work is implemented in the Python package STACIE,
which we developed and made publicly available on
PyPI (pip install stacie),
Conda Forge (conda install -c conda-forge stacie) and
GitHub (\url{https://github.com/molmod/stacie}),
along with extensive documentation.
The source code has been deposited on Zenodo (\url{https://doi.org/10.5281/zenodo.18077751}).

The ACID dataset used to validate the accuracy of STACIE's implementation of the Lorentz model
is available at \url{https://doi.org/10.5281/zenodo.20026468}.
Molecular dynamics trajectory data used in this study are available at
\url{https://doi.org/10.5281/zenodo.15689551}
as npz files corresponding to each simulation condition.

\section*{Supporting Information} \label{sec:supporting}
The Supporting Information is available free of charge at \url{https://pubs.acs.org/doi/10.1234/acs.abcd.1234}. \\

\begin{itemize}
  \item
  PDF document with additional display items and background information:
  \begin{itemize}
    \item Section~S1: Validation of STACIE's Lorentz model using the ACID test set
    \item Section~S2: Derivation of the five uncorrelated deviatoric pressure components for viscosity calculations
    \item Section~S3: Literature survey of transformations of the pressure tensor used in EMD-based shear viscosity calculations
    \item Section~S4: Analysis of the McEwen--Paluch model for the viscosity of 2,2,4-trimethylhexane
    \item Section~S5: STACIE \& TDM shear viscosity results for all pressures (full trajectories)
    \item Section~S6: STACIE \& TDM shear viscosity results for truncated trajectories at \SI{500}{\mega\pascal}
    \item Section~S7: STACIE \& TDM shear viscosity results for truncated trajectories at \SI{1000}{\mega\pascal}
    \item Section~S8: Analysis of the five uncorrelated deviatoric pressure components for viscosity calculations
  \end{itemize}
  \item
  ZIP file with CSV files containing the numerical data of all figures and tables in the main text.
\end{itemize}

\section*{Author Information}\label{sec:authorinfo}
\subsection*{Author Contributions}
G.T. performed and analyzed simulations, designed and validated algorithms and contributed to software development.
D.F. and T.V. conceptualized and supervised the project.
T.V. designed and validated algorithms, and led software development.
All authors wrote the original draft of the manuscript.

\subsection*{Notes}
The authors declare no competing financial interest.

\section*{Acknowledgments} \label{sec:acknowledgments}
This research was funded by the Research Board of Ghent University with grant number BOF/24J/2021/118.
Computational resources were provided by Ghent University's Stevin High-Performance Computing (HPC) infrastructure and the Flemish Supercomputer Center (VSC), funded by the Research Foundation Flanders (FWO).

\bibliography{references}

\includepdf[pages=-]{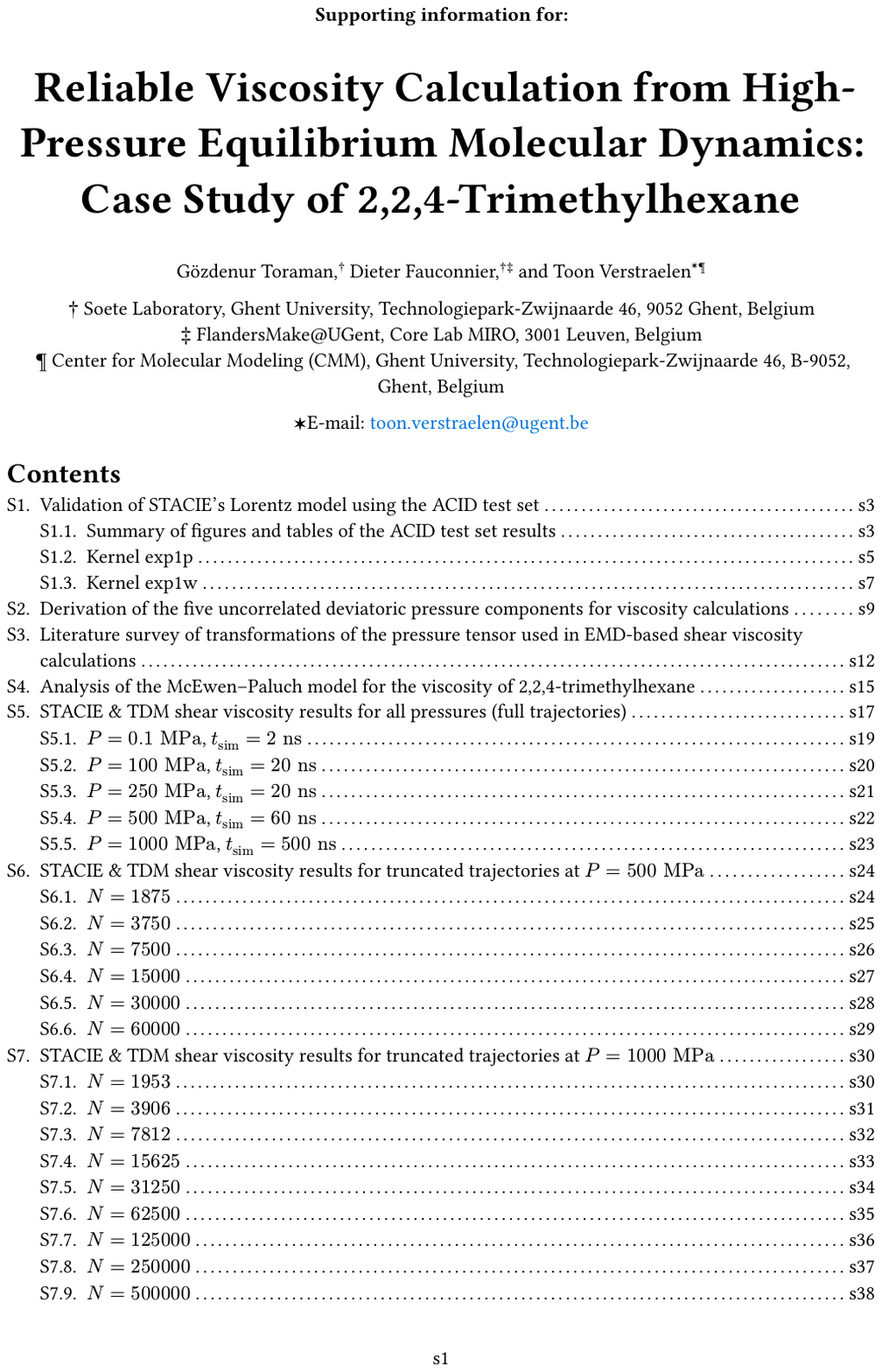}
\end{document}